# Strong Neel ordering and luminescence correlation in a two-dimensional antiferromagnet

*Yongheng Zhou, Kaiyue He, Huamin Hu, Gang Ouyang, Chao Zhu, Wei Wang, Sichen Qin, Ye Tao, Runfeng Chen, Le Zhang, Run Shi, Chun Cheng, Han Wang, Yanjun Liu, Zheng Liu, Taihong Wang[\*], Wei Huang, Lin Wang[\*], Xiaolong Chen[\*]*

Y. Zhou, L. Zhang, H. Wang, Prof. Y. Liu, Prof. T. Wang, Prof. X. Chen
Department of Electrical and Electronic Engineering, Southern University of Science and Technology, Shenzhen 518055, China
E-mail: chenxl@sustech.edu.cn; wangth@sustech.edu.cn

K. He, Prof. W. Wang, Prof. W. Huang, Prof. L. Wang
Key Laboratory of Flexible Electronics (KLOFE) & Institute of Advanced Materials (IAM), Jiangsu National Synergetic Innovation Center for Advanced Materials (SICAM), Nanjing Tech University (Nanjing Tech), 30 South Puzhu Road, Nanjing 211816, China
E-mail: iamlwang@njtech.edu.cn

H. Hu, Prof. G. Ouyang
Key Laboratory of Low-Dimensional Quantum Structures and Quantum Control of Ministry of Education, and Key Laboratory for Matter Microstructure and Function of Hunan Province, Hunan Normal University, Changsha 410081, China

Dr. C. Zhu
SEU-FEI Nano-Pico Center, Key Laboratory of MEMS of Ministry of Education, Collaborative Innovation Center for Micro/Nano Fabrication, Device and System, Southeast University, Nanjing 210096, China

Dr. C. Zhu, Prof. Z. Liu
School of Materials Science and Engineering Nanyang Technological University, Singapore 639798, Singapore

S. Qin, Prof. W. Huang
Frontiers Science Center for Flexible Electronics (FSCFE), Shaanxi Institute of Flexible Electronics (SIFE) & Shaanxi Institute of Biomedical Materials and Engineering (SIBME), Northwestern Polytechnical University (NPU), 127 West Youyi Road, Xi'an 710072, China

Prof. Y. Tao, Prof. R. Chen
State Key Laboratory of Organic Electronics and Information Displays & Institute of Advanced Materials (IAM), Nanjing University of Posts & Telecommunications, 9 Wenyuan Road, Nanjing 210023, China

Dr. R. Shi, Prof. C. Cheng
Department of Materials Science and Engineering, Southern University of Science and Technology, Shenzhen 518055, China

**Abstract**

Magneto-optical effect has been widely used in light modulation, optical sensing and information storage. Recently discovered two-dimensional (2D) van der Waals layered magnets are considered as promising platforms for investigating novel magneto-optical phenomena and devices, due to the long-range magnetic ordering down to atomically-thin thickness, rich species and tunable properties. However, majority 2D antiferromagnets suffer from low luminescence efficiency which hinders their magneto-optical investigations and applications. Here, we uncover strong light-magnetic ordering interactions in 2D antiferromagnetic $MnPS_3$ utilizing a newly-emerged near-infrared photoluminescence (PL) mode far below its intrinsic bandgap. This ingap PL mode shows strong correlation with the Neel ordering and persists down to monolayer thickness. Combining the DFT, STEM and XPS, we illustrate the origin of the PL mode and its correlation with Neel ordering, which can be attributed to the oxygen ion-mediated states. Moreover, the PL strength can be further tuned and enhanced using ultraviolet-ozone treatment. Our studies offer an effective approach to investigate light-magnetic ordering interactions in 2D antiferromagnetic semiconductors.

**1. Introduction**

The emerging atomically-thin van der Waals (vdW) layered magnet has offered an exciting platform for investigating light-matter interactions down to monolayer limit.[1-13] There has been significant progress in revealing the magnetic properties of vdW layered magnets through optical approaches in recent years, such as the thickness-dependent magneto-optical Kerr effect and magneto-optical Raman effect in $CrI_3$[2, 14-16] and $Cr_2Ge_2Te_6$,[3] correlation between helical luminescence and the ferromagnetic ordering in monolayer $CrI_3$,[17] coherent many-body exciton and observation of spin-correlated linearly polarized emission in $NiPS_3$,[4, 5, 9] Fano resonance induced by quantum interference in $CrPS_4$[18] and exciton-magnon coupling at MoSe$_2$/MnPSe$_3$ interfaces.[19] Among various vdW layered magnets, the wide bandgap (~ 2.9 eV) MnPS$_3$ hosts a robust long-range Heisenberg-type antiferromagnetic ordering down to atomically-thin thickness.[20-23] Its Neel transition temperature ($T_N$) ~ 78 K is almost independent of thickness, which has been confirmed by Raman spectroscopy and tunnel transport measurements.[24-29] Besides, few-layer MnPS$_3$ exhibits a linear magnetoelectric phase below Neel temperature due to the breaking of spatial-inversion and time-reversal symmetry.[30, 31] Despite the much progress on magnetic properties of MnPS$_3$,[20-32] the investigation on the light emission property and its correlation with antiferromagnetic ordering were scarce.

In this work, we investigate the strong interactions between the antiferromagnetic ordering and the light emission properties in 2D MnPS$_3$. We uncover a new photoluminescence (PL) mode, far below its intrinsic electronic bandgap ~ 2.9 eV.[21] Both PL intensity and energy show strong correlation with the antiferromagnetic ordering. More importantly, this correlation can still persist when MnPS$_3$ thickness approaches the 2D limit (monolayer and bilayer thickness). Density-functional theory (DFT) calculations, X-ray photoelectron spectroscopy (XPS) and scanning transmission electron microscope (STEM) suggests this unusual near-infrared PL mode originates from ingap electron transitions assisted by chemically absorbed oxygen element with spin configuration coupled to antiferromagnetic ordering of Mn ions. At last, we show an effective approach to artificially introducing oxygen absorption on MnPS$_3$ for tuning and enhancing the strength of the ingap PL mode.

## 2. Results

### 2.1 Characterization of MnPS$_3$

**Figure 1**a shows the atomic vdW structure of MnPS$_3$, belonging to a monoclinic system with space group of C2/m. Each unit cell of monolayer MnPS$_3$ is formed by two Mn$^{2+}$ ions and one P$_2$S$_6^{4-}$ cluster. Mn$^{2+}$ ions form a honeycomb structure and each Mn$^{2+}$ ion is surrounded by six S atoms. The large field of view annular dark-field (ADF)-STEM image (left panel of Figure 1c) along [103] zone axis (top view) suggests the high quality of MnPS$_3$ crystals, where the sharp spots in fast Fourier transform (top-right panel of Figure 1c) fit well with the simulated one (bottom-right panel of Figure 1c), including the strong diffracted ($33\bar{1}$) and (060) facets as well as the weak diffracted (020) and (040) facets. The high resolution ADF-STEM image (left panel of Figure 1d) shows the in-plane six-fold atom symmetry of MnPS$_3$. While the intensity profile of atom chain highlighted with the yellow rectangle reveals the out-of-plane stacking feature that Mn and S atom columns possess higher atomic number *Z*-contrast than that of P and S atom columns (right panel of Figure 1d). Due to the weak vdW interactions between each layer, few-layer MnPS$_3$ can be obtained using the mechanical exfoliation method (also known as "the Scotch Tape Method").[7] Here, we deposited few-layer MnPS$_3$ on several kinds of substrates, including 300 nm-SiO$_2$/Si wafers, gold-covered SiO$_2$/Si wafers and PDMS films. Among them, gold-covered SiO$_2$/Si wafers show the weakest background signal and show insignificant influence to the PL properties of atomically thin MnPS$_3$ (see **Supplementary Figure S1**). Thus, gold-covered SiO$_2$/Si wafers were chosen as substrates for the PL investigation. The thickness of MnPS$_3$ was confirmed by the atomic force microscope and the optical contrast measurement. The optical contrast is in a good linear relation with the thickness, which shows a 0.7% step of optical contrast when adding/removing one MnPS$_3$ layer (see **Supplementary Figure S2**). We further performed the Raman spectroscopy of MnPS$_3$ flakes

at room temperature. As shown in Figure 1b, eight Raman modes ($P_1$-$P_8$) are all detected and their positions correspond well with previous reported values.[24, 26, 27, 33] The high frequency modes ($P_3$-$P_8$) are ascribed to the molecular-like vibrations of $(P_2S_6)^{4-}$ clusters, and the low frequency modes ($P_1$-$P_2$) are from the vibration of $Mn^{2+}$ ions.

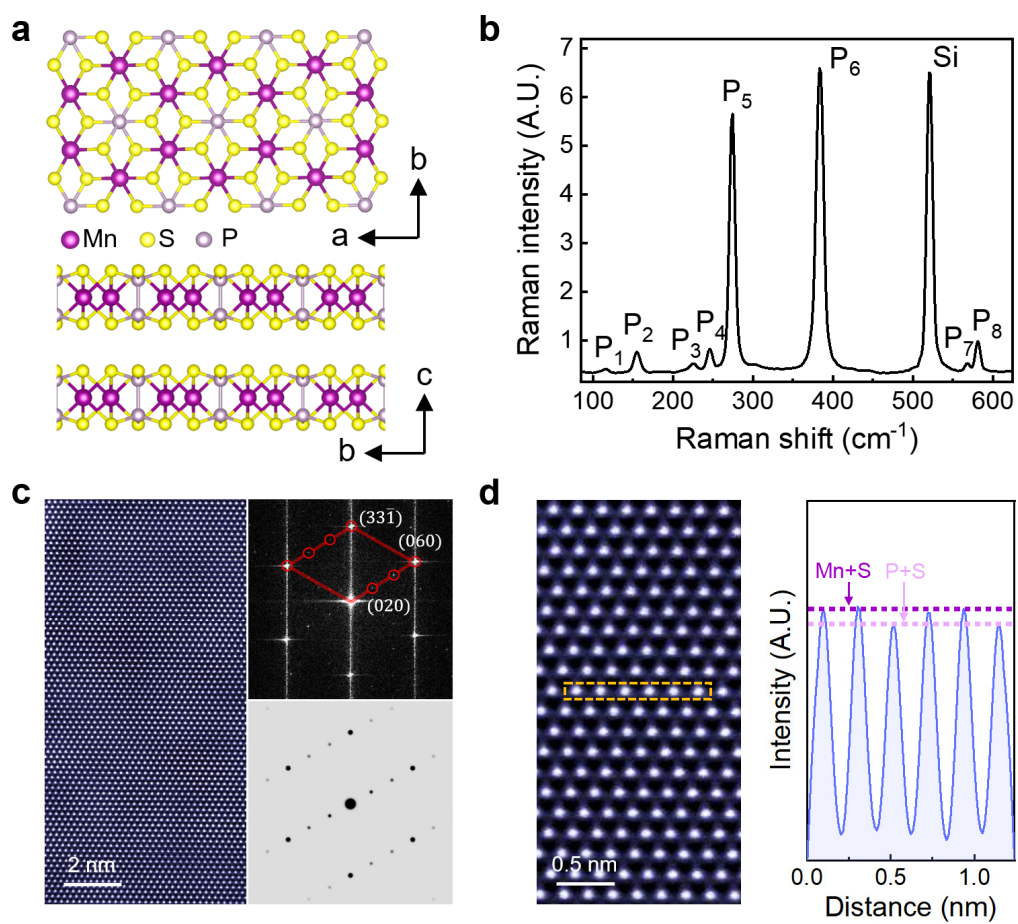

**Figure 1.** Crystalline structure, Raman spectrum and TEM images of $MnPS_3$. a) Schematic side and top views of vdW layered $MnPS_3$ crystal. b) Raman spectrum of bulk $MnPS_3$ measured at room temperature. c) Large field of view ADF-STEM image (left panel), fast Fourier transform (top-right panel) and simulated diffraction pattern (top-right panel) of $MnPS_3$. d) Atomic resolution ADF-STEM image (left panel) and the intensity profile of the rectangular region (right panel) of $MnPS_3$.

## 2.2 Photoluminescence in MnPS$_3$

To investigate the interactions between light emission and antiferromagnetic ordering in MnPS$_3$, we first performed temperature-dependent PL measurements. We excited MnPS$_3$ flakes with 405, 532 and 638 nm lasers. Under the illumination with all these three excitation wavelengths, a near-infrared PL peak P$_{nir}$ (~ 940 nm/1.32 eV) can be revealed, far below its intrinsic electronic bandgap (~ 2.9 eV).[21] **Figure 2**a shows the PL spectra of a 46-layer MnPS$_3$ flake under 532 nm laser excitation from 10 to 290 K. The near-infrared PL peaks show very good Gaussian shape and can be well fitted using a single Gaussian peak. Besides, the PL peaks can also be fitted using two Gaussian peaks P1 and P2. Both of them shows a very similar trend of temperature-dependent PL intensity (see **Supplementary Figure S3**). However, P2 shows negligible PL intensity in the temperature range from 10 to 180 K, and it could hardly influence the shape of the PL peak. Hence, the single-Gaussian-peak fitting is chosen to describe PL properties of near-infrared PL peaks. At 10 K, the PL peak is located at 941 nm (~1.32 eV) with a full-width-at-half-maximum (FWHM) of ~111 nm. The energy of PL peak shows clear signature of the antiferromagnetic ordering transition (see Figure 2b). When temperature ranges from 10 to 78 K, the peak energy shows a shift of 6 meV with a linear slope of 0.098 meV K$^{-1}$. Above the transition temperature, the peak energy shows a shift of 34 meV from 78 K to 200 K with a steeper slope of 0.290 meV K$^{-1}$. The changes in photon energy below the transition temperature might be the signature of magnon-exciton interactions according to previous reports.[34-37] The red solid line is the linear fitting curve using data from 90 to 200 K and is prolonged to 10 K, which is denoted as $E_{LF}$. The energy difference ($\Delta E$) between the experimental data ($E_{Exp}$) and $E_{LF}$ is extracted by the equation $\Delta E = E_{Exp} - E_{LF}$. As shown in Figure 2c, $\Delta E$ vanishes when temperature is above the Neel transition temperature $T_N$ ~ 78 K of MnPS$_3$.[22, 24-28] When temperature is below $T_N$, $\Delta E$ is between 0 to 14 meV which is at the same order with previous reported value of magnons.[38] This suggests that magnon-exciton interactions might exist in MnPS$_3$. Interestingly, the near-infrared PL peak position shows blueshift when temperature increases, which is opposite to those of majority 2D materials. One possible explanation is the local-strain-induced blueshift of PL in MnPS$_3$. Previous study of PL in thin-film black phosphorus shows that the PL blueshift with increasing temperature is due to the large thermal strain in black phosphorus.[39] In analogy, since the infrared PL in MnPS$_3$ originates from the oxygen-bonded defects (which will be discussed later), the local thermal properties in the defect region might be different from those in the bulk region. We guess that the defect region might experience additional strain when temperature changes which leads to the unusual PL blueshift. However, this is only one possible scenario. The physical mechanism behind this is still ambiguous and needs further investigation to fully understand this anomalous phenomenon.

Temperature-dependent PL intensity of MnPS$_3$ shows a peak shape (see Figure 2d), which is different from that of paramagnetic semiconductors. In most of paramagnetic semiconductors, the PL intensity monotonously decreases as the temperature increases due to the enhancement of phonon-exciton interactions at high temperatures, and the PL position shows redshift following the conventional Varshni equation. However, when temperature increases, the PL intensity of MnPS$_3$ gradually increases and reaches the maximum value near $T_N$, while the PL intensity becomes weaker at higher temperatures. This suggests interactions between light emission and antiferromagnetic ordering might exist in MnPS$_3$. To extract transition temperatures from temperature-dependent PL spectra, the slope of intensity-temperature curve is calculated and shown in Figure 2e. The slope-temperature curve shows a linear region from 60 to 90 K. The linear fitting can be used to extract the maximum-PL-intensity temperature when the slope is zero, which is T= 79.8±7.8 K. This value agrees well with the Neel transition temperature $T_N$ ~78 K of bulk MnPS$_3$. We also noticed that the PL intensity-temperature crossover is very broad from 60 to 90K. Similar broad crossover feature can also be observed in the temperature-dependent susceptibility measurement of bulk MnPS$_3$ (see **Supplementary Figure S4**). Since the antiferromagnetic to paramagnetic ordering transition is not a transient process, short-range spin-spin correlation will be preserved in a temperature range above $T_N$. Hence, this broad crossover feature of temperature-dependent PL is expected in antiferromagnetic MnPS$_3$. The evolution of FWHM of the PL peak with temperature shows that the FWHM broadens at higher temperatures with no apparent signature of antiferromagnetic ordering transitions (see **Supplementary Figure S5**).

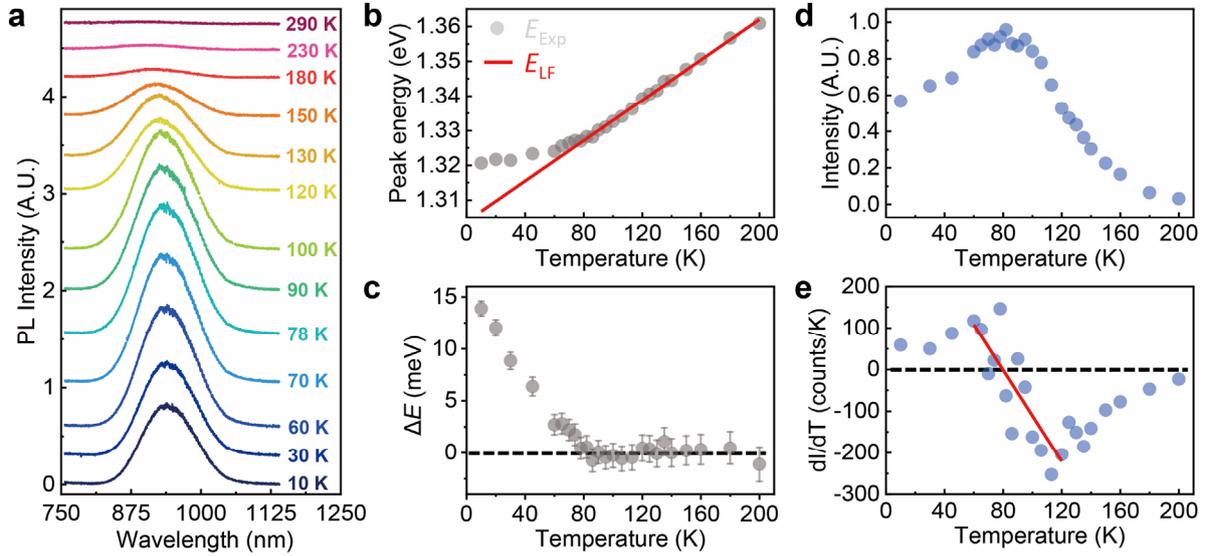

**Figure 2.** Temperature-dependent PL spectra of a 46-layer MnPS$_3$. a) PL spectra of 46-layer MnPS$_3$ sample at different temperatures excited by a 532 nm laser. The laser power is 100 μW. b) Peak energy as function of temperatures. The scatter is experimental data (denoted as $E_{Exp}$), while the red solid line ($E_{LF}$) is the linear fitting curve using data from 90 to 200 K and is prolonged to 10 K. c) Energy difference ($\Delta E$) between experimental and linear fitting data as function of temperatures. d) Temperature-dependent PL intensity. e) The slope of PL-intensity-temperature curve. The red solid line is the linear fitting curve in the temperature range from 60 to 90 K.

Then we explored the thickness-dependent PL properties of MnPS$_3$. **Figure 3**a shows the PL spectra of 1-layer, 2-layer, 4-layer, 6-layer, 13-layer, 26-layer and 46-layer MnPS$_3$ samples at temperature of 78 K and with incident power of 100 μW (532 nm laser source). The PL intensity decreases dramatically as thickness reduces from bulk to few layers (see Figure 3b). For 1 to 3-layer MnPS$_3$ samples, their PL intensities are too weak to observe under 532 nm laser excitation, but can be revealed by 405 nm laser excitations (see Figure 3a and **Supplementary Figure S6** and **Figure S7**). On the other hand, the peak position is almost independent of thickness (see Figure 3c), indicating the weak layer coupling in MnPS$_3$. The temperature-dependent PL intensities for different thickness are summarized in Figure 3d. The maximum PL intensities and crossover peak positions for few-layer (> 3 layers) and bulk samples all occur near 78 K. This thickness-independent antiferromagnetic ordering transition revealed by our PL spectra (see Figure 3e) is consistent with previous results using Raman and tunnel transport measurements.[22, 24, 26, 40] Correlation between PL properties (intensity and position) and antiferromagnetic ordering transitions can also be observed in 1-layer and 2-layer samples under 405 nm laser excitation (see Figure 3a and Supplementary Figure S6 and S7). A very recent theoretical study predicts that the antiferromagnetic order of MnPS$_3$ can persist down to monolayer,[41] which provides important theoretical supports to out experimental results. The extracted transition temperature of 1-layer and 2-layer samples were around 63.4±2.4 K and 74.5±5.0 K, respectively. Although both 1-layer and 2-layer samples seem to show smaller average transition temperatures than those of bulk samples, given the low PL intensity and large signal fluctuations, further experiments are needed to confirm the observation. The PL spatial variations of 1-layer and 2-layer samples are also studied (see **Supplementary Figure S8**). Temperature-dependent PL measurements at different positions of 1-layer and 2-layer samples show that the extracted transition temperatures at different positions show some variation, but the variation is within the range of experimental error. The variation is random and show no obvious correlation to spatial positions (see **Supplementary Figure S9** to **Supplementary Figure S11**).

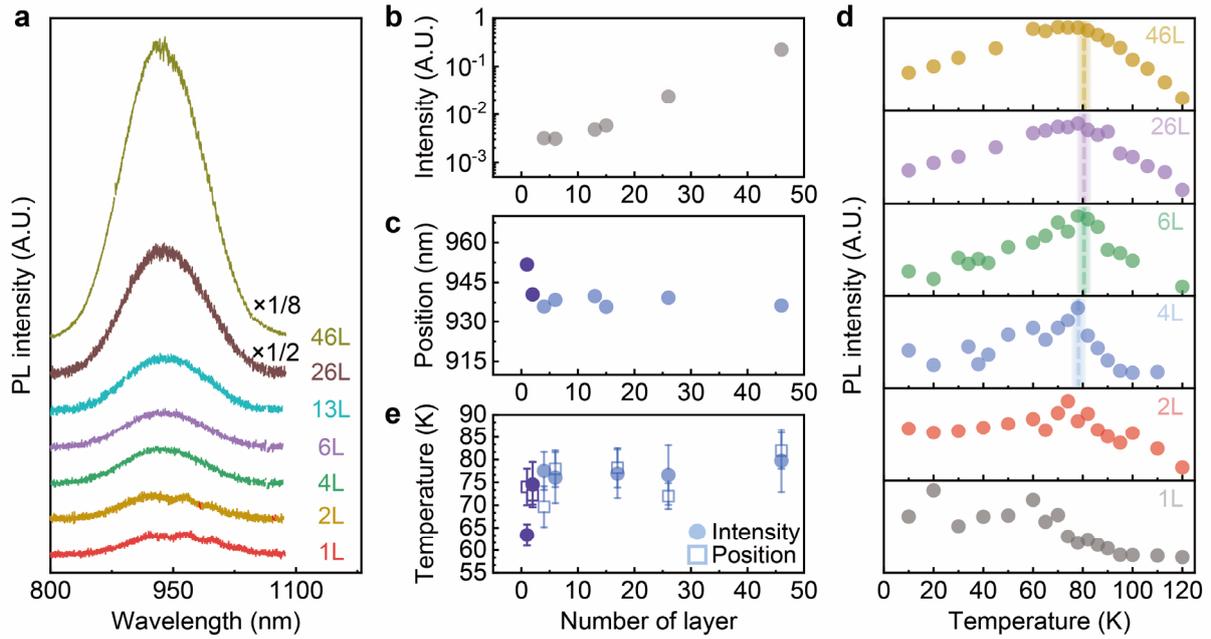

**Figure 3.** Thickness-dependent PL spectra of MnPS$_3$. a) PL intensity of MnPS$_3$ with various thickness at 78 K. The excitation laser wavelength is 532 nm for 4-layer to 46-layer samples. The PL of 1-layer and 2-layer samples were obtained under 405 nm excitation laser with power of 500 μW. b,c) PL intensity (b) and peak position (c) as function of thickness at 78 K. The laser power is 100 μW. d) Temperature-dependent PL intensities of MnPS$_3$ with various thickness. The dashed lines show the temperature of maximum PL intensity. e) Transition temperatures extracted from temperature-dependent PL intensities and positions as function of layer numbers. The dots and boxes represent transition temperatures extracted from intensity and peak position, respectively. The purple dots and boxes in (c) and (e) indicate the transition temperature of 1-layer and 2-layer samples, respectively, measured by the 405 nm laser with power of 500 μW.

For 405 nm laser excitation with photon energy larger than the intrinsic bandgap of MnPS$_3$, we observe a new visible PL peak P$_{vis}$ at ~ 566 nm (2.20 eV) for the 46-layer sample (see **Figure 4**a), as not resolved under 532 and 638 nm laser excitations. The PL spectra under 405 nm laser show additional periodic peaks, which will shift and become less obvious when the optical filter is changed. This indicates those periodic peaks does not come from MnPS$_3$, but from the light interference in our PL systems. Similar antiferromagnetic ordering transition phenomena are observed for the P$_{nir}$ with a maximum PL intensity at 78 K under 405 nm laser excitations (see Figure 4b). For the P$_{vis}$, its PL intensity reaches the minimum value at 78 K, showing an inverse temperature-dependent trend compared with that of the P$_{nir}$ peak. This might indicate that there exists a competition effect of electron transitions between $P_{vis}$ and $P_{nir}$ peaks. As shown in Figure 4c, the PL peak position of $P_{vis}$ shows conventional redshift as the temperature increases in low temperature (<100 K) and high temperature ranges (>150 K). However, an anomalous blueshift in temperature range from 100 to 150 K is observed. This phenomenon is further confirmed in other MnPS$_3$ samples (see **Supplementary Figure S12**). A possible scenario for this anomalous blueshift near 120 K might be related to the magnetic phase transition. According to previous reports,[42, 43] MnPS$_3$ has a second magnetic phase above $T_N$, where the short-range spin-spin correlation survives until the temperature rises above 120 K. Previous reports on temperature-dependent Raman spectra showed the existence of interactions between magnetic ordering and phonons.[24] Hence, it is possible that the exciton-phonon coupling might affect the exciton energy and lead to the anomalous blueshift of the visible PL peak position. Further studies are needed to fully understand this anomalous phenomenon near 120 K. Besides, the influence of different excitation wavelengths and powers on PL properties are studied. First, temperature-dependent PL properties of MnPS$_3$ flake were investigated using 405, 532 and 638 nm lasers, respectively. The extracted transition temperatures under different lasers show no obvious temperature dependence (see **Supplementary Figure S13** and **Table S1**). In addition, the power-dependent PL study with excitation laser power ranging from 100 to 300 μW was performed and the extracted transition temperatures also show no clear power dependence (see **Supplementary Figure S14**). However, the excitation power could possibly affect the transition temperature if the power is high enough to heat the samples. Additionally, in order to provide useful information for the origin of visible and near-infrared PL peaks, the power-dependent PL intensities were studied at room temperature with higher excitation laser power up to 1300 μW as shown in Figure 4d. For the near-infrared peak, the PL intensity deviates from the linear power dependence and starts to show saturation when power increases above 800 μW, while for the visible PL peak, the intensity shows linear power dependence. These results indicate that the near-infrared and visible PL peaks have different origin. The saturation characteristics of the near-infrared PL peak suggests that it might originate from the defect-related transitions, which is consistent with our theoretical results shown below.

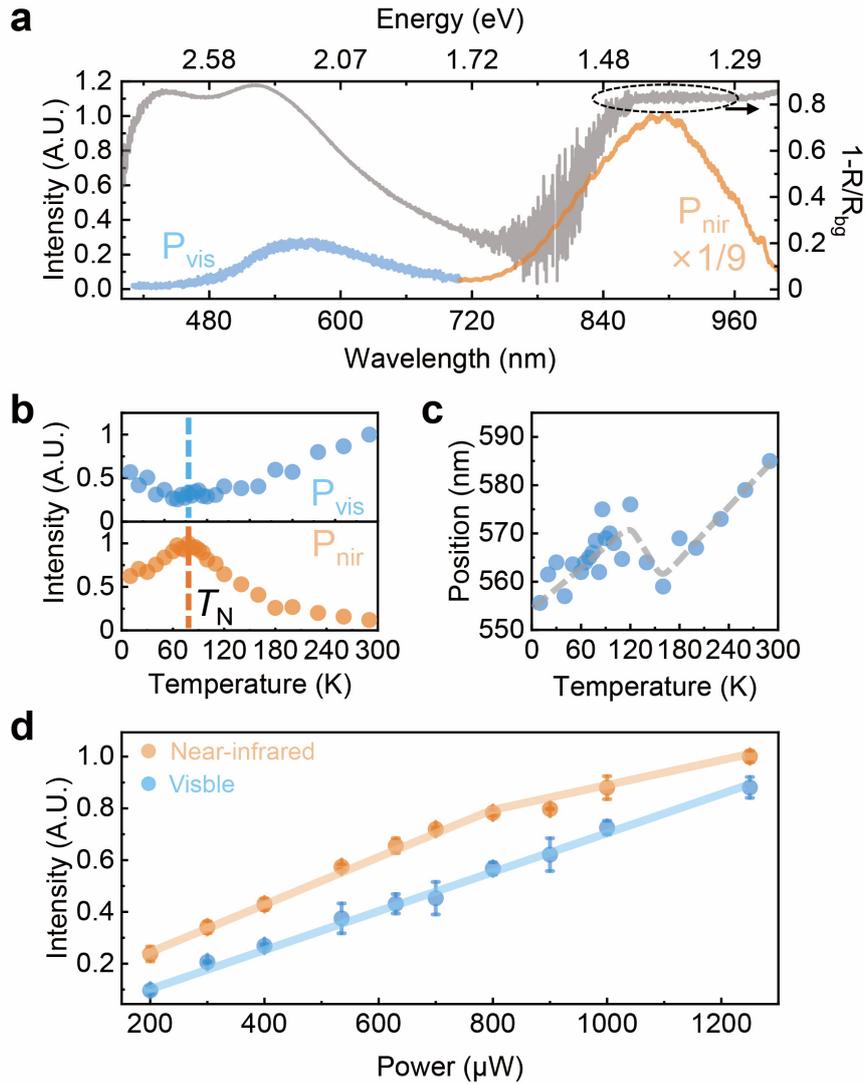

**Figure 4.** PL spectra of MnPS$_3$ under 405 nm laser excitation. a) PL spectra of 46-layer MnPS$_3$ sample at 78 K excited by a 405 nm laser. Gray line shows the reflection spectrum of MnPS$_3$ sample. b) PL intensity of P$_{vis}$ (top panel) and P$_{nir}$ (lower panel) as function of temperatures. Dashed line shows the Neel transition temperature ~78 K. c) Temperature-dependent position of the visible PL peak. The gray dashed line serves as the guideline. d) Power-dependent PL intensities (shown by dots) at room temperature with laser power up to 1300 μW. The blue and orange solid lines servers as guidelines.

## 2.3 Origin of the photoluminescence peaks

The reflection spectra of MnPS$_3$ show three pronounced absorption peaks (438, 521 and 870 nm) in the range from 400 to 1000 nm (see Figure 4a). The 438 nm (2.84 eV) peak is consistent with reported transmission spectra,[21, 44] which can be attributed to the electron transitions between conductance to valence bands. DFT calculations (using PBE+U method) show that monolayer and bulk MnPS$_3$ with antiferromagnetic phase are both direct bandgap semiconductor (see **Figure 5**a,b). Besides, the bandgap size of MnPS$_3$ weakly depends on thickness, indicating a weak layer coupling in MnPS$_3$. As shown in Figure 5c, the bandgap size is ~ 2.50 eV in monolayer MnPS$_3$ and decreases to 2.30 eV in bulk form. The calculated bandgap size is smaller than the value (~ 2.84 eV) obtained from experimental transmission/reflection spectra. This is reasonable since the PBE method always underestimates the bandgap size. Nevertheless, our calculated bandgap size shows good agreement with previous published theoretical results and can qualitatively explain our experimental observations.[45]

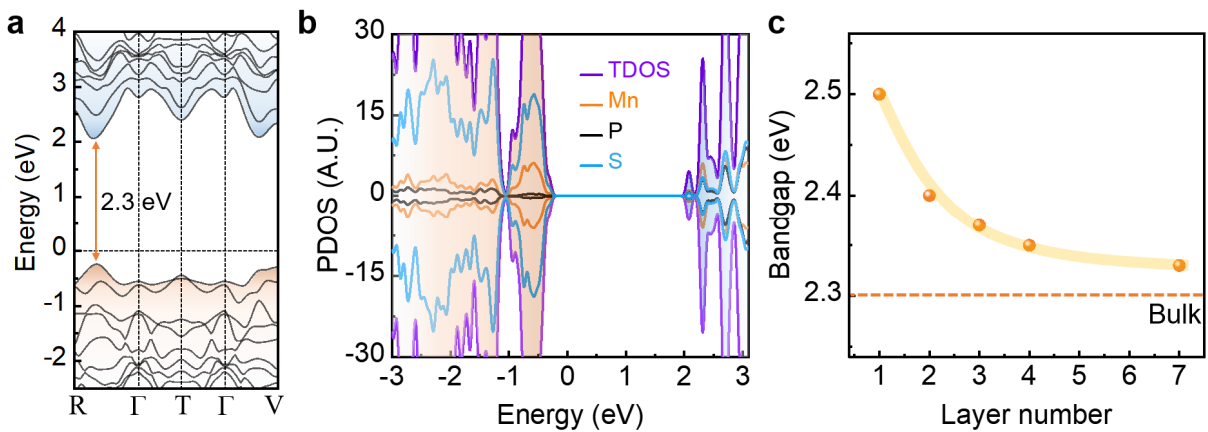

**Figure 5.** DFT calculations of pristine MnPS$_3$. a) Band structure of bulk MnPS$_3$ crystal with AFM ordering. b) Projected density of states (PDOS) of bulk MnPS$_3$. The Fermi level is set to zero. c) Calculated bandgap size (scatters) as a function of thickness. The dashed line indicates the bandgap size of bulk MnPS$_3$. The yellow solid line serves as the guideline.

The exciton binding energy in few-layer and bulk MnPS$_3$ is at order of 100 meV according to previous theoretical calculations[46]. As a result, the P$_{vis}$ (~ 2.20 eV) and P$_{nir}$ (~ 1.35 eV) photons are unlikely due to the electron transitions between conductance and valence bands. From the reflection spectrum (see Figure 4a), the 521 and 870 nm absorption peaks should correspond to P$_{vis}$ and P$_{nir}$ photons. This indicates that P$_{vis}$ and P$_{nir}$ photons might originate from electron transitions between midgap and conductance/valence bands. X-ray photoelectron spectroscopy (XPS) shows two pronounced emission peaks near 530 eV, indicating the chemical absorption of oxygen element in MnPS$_3$ (see **Figure 6**a). According to previous XPS research on MnO$_2$ and Na$_{0.7}$MnO$_{2.05}$, the 529.8 and 531.4 eV peaks can be attributed to the Mn-O-Mn and Mn-O-H bonds, respectively.[47-49] The existence of oxygen element in MnPS$_3$ is further confirmed by energy-dispersive X-ray spectroscopy (EDS) analysis using STEM (see Figure 6b). The spectrum line indicates an obvious O Kα peak at 0.525 keV, and like Mn, P and S elements the inserted EDS mapping present the uniform distribution of O with the atom ratio of 2.1%. Then we construct a theoretical model (2-layer MnPS$_3$ as an example) to verify the formation of midgap band in MnPS$_3$ with presence of Mn-O-Mn and Mn-O-H bonds (see Figure 6c). The calculated projected density of states (PDOS) with presence O and H ions exhibits two new impurity energy levels inside bandgap, I$_1$ (spin up) and I$_2$ (spin down) (see Figure 6d). I$_2$ energy level has more holes because it is located on the right side of the Fermi level, which indicates that the electrons in conduction band minimum (CBM) are easier to transit to I$_2$ level. Moreover, by analyzing the electron spin density of MnPS$_3$ with presence O and H ions (see **Supplementary Figure S15**), we find that the vicinity of the O adsorption site is mainly contributed by the spin-down electrons, which implies that the spin-down electrons dominate during the energy level transition.

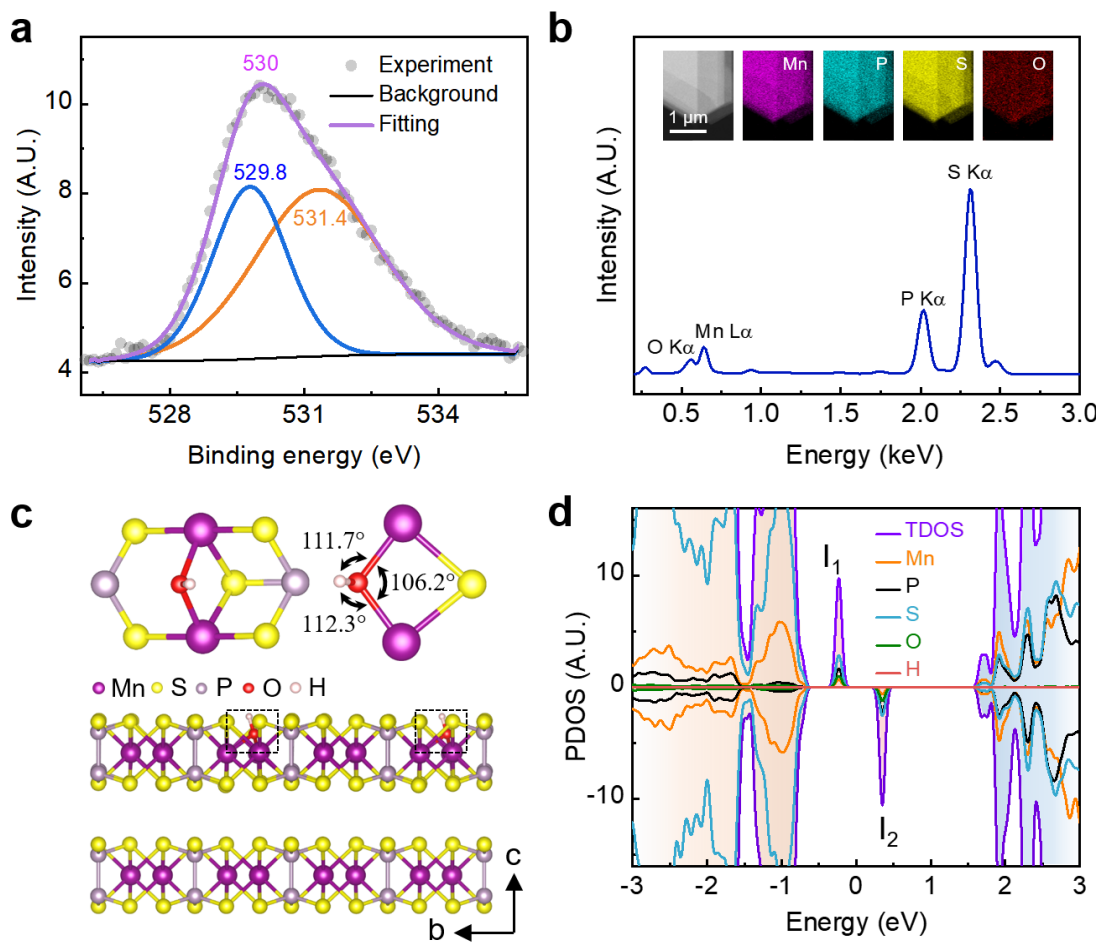

**Figure 6.** Origin of PL peaks in MnPS$_3$. a) XPS of bulk MnPS$_3$ indicates the chemical absorption of O element. b) EDS spectrum and mapping confirm the absorption of O element in MnPS$_3$. c) DFT model of a bilayer MnPS$_3$ with O and H ions. d) Calculated PDOS of a bilayer MnPS$_3$ with O and H ions. The Fermi level is set to zero.

We also considered other possible scenarios, such as S or Mn vacancy induced impurity bands. However, our calculation results show that both S and Mn vacancies cannot induce midgap states inside bandgap (see **Supplementary Figure S16**). This indicates that although S and Mn vacancies might exist in MnPS$_3$, they are unable to contribute to P$_{nir}$ and P$_{vis}$ photons.

Oxygen absorption is further intentionally introduced into MnPS$_3$ using ultraviolet-ozone (UVO) treatment, which is widely used to introduce oxygen bonds in 2D materials.[50-55] Characterization result shows that the morphology and crystal structure of the sample didn't change after 7 min UVO treatment (see **Supplementary Figure S17**). In **Figure 7a**, we compare the PL spectra for the MnPS$_3$ flake under UVO treatment for 0, 3 and 7 minutes, respectively. The as-exfoliated sample shows a PL position of ~888 nm at room temperature. After UVO treatment, the peak position remains the same, while the intensity is significantly enhanced. For longer treatment duration, the PL intensity saturates. In addition, the PL spectra before and after the UV radiation in vacuum are shown in the Figure 7b. The PL behavior of the sample shows no difference after the UV radiation in vacuum environment, which indicates that the UV radiation itself will not affect the PL behavior of MnPS$_3$ unless oxygen molecules take part in the process. This result further confirms the origin of this ingap photon emission mode in MnPS$_3$.

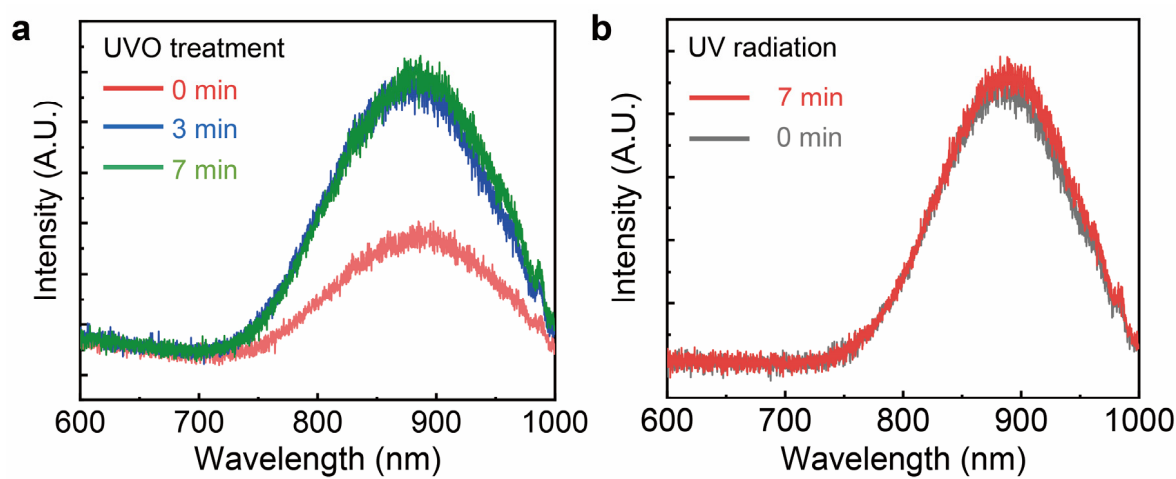

**Figure 7.** PL spectra of MnPS$_3$ under the ultraviolet-ozone (UVO) treatment. a) PL spectra of MnPS$_3$ under UVO treatment for 0, 3 and 7 mins, respectively. Pronounced enhancement of PL intensity is observed. b) The PL spectra before and after 7 min UV radiation in vacuum. The power of the UV radiation is 100 mW. The excitation laser wavelength is 405 nm and the power density of the laser is 1 μW /cm$^2$.

At last, we systemically investigated the influence of air exposure to the PL properties of atomically thin MnPS$_3$ samples. A 1-layer sample was prepared and measured in vacuum. Then it was exposed to air for 60 minutes. The PL intensity of the sample shows no obvious decrease after being exposed to air for 10 minutes. As the exposure time further increases to 40 minutes, the PL intensity starts to show some decrease (see **Supplementary Figure S18**a). PL positions of the sample were also extracted and show no clear dependence to the exposure time (see Supplementary Figure S18b). This indicates that MnPS$_3$ has a relatively good air stability. According to our experimental condition, the exposure time (within 1 minute) is not long enough to have significant influence on the PL properties of the 1-layer MnPS$_3$ sample. Hence, it will have even smaller influence on 2-layer and thicker samples.

## 3. Discussion and conclusion

The near-infrared peak P$_{nir}$ is mainly contributed by the electron transitions at O ion site in antiferromagnetic MnPS$_3$. Both magnons and phonons can interact with excitons and affect the process of electron transitions. **Figure 8** shows the schematic diagram of the temperature-dependent PL intensity of P$_{nir}$. The purple and blue curves represent the PL intensity contributed by phonon-exciton and magnon-exciton interactions, respectively. 1) At low temperature (< $T_N$), phonon density is low and the phonon-exciton interaction is weak. Hence, magnon-exciton interactions have a stronger impact on the electron-hole recombination process than phonon-exciton interactions. According to our theoretical calculations, the long-range antiferromagnetic ordering only sustains spin down electrons at O ion site which lowers the electron-hole recombination probability. Hence, lower temperature leads to lower electron density near O ion and lower PL intensities. 2) At high temperature (> $T_N$), the long-range antiferromagnetic ordering will gradually be destroyed by the large thermal fluctuation energies. Hence magnon-exciton interactions will gradually vanish and phonon-exciton interactions start to significantly affect the electron-hole recombination process. The stronger nonradiative scatterings at higher temperatures will lead to lower PL intensities. As a result, the total PL intensities (denoted by the red line in Figure 8) show a peak shape, originating from the competition between magnon-exciton and phonon-exciton interactions.

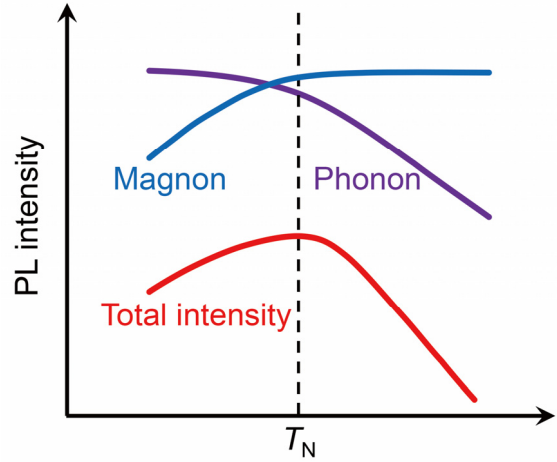

**Figure 8.** Schematic diagram of the temperature-dependent PL intensity of $P_{nir}$. The purple and the blue curves represent the PL intensity contributed by phonon-exciton and magnon-exciton interactions, respectively. The total PL intensity is denoted by the red line.

Since there is competition between $P_{nir}$ and $P_{vis}$ photons, $P_{vis}$ intensity shows a reversed temperature dependence compared with that of $P_{nir}$. Our explanation is only one possible scenario and further theoretical studies on the Neel-transition-dependent PL of $MnPS_3$ are encouraged. Nevertheless, we have demonstrated strong interactions between the Neeling ordering and light emission properties in 2D antiferromagnetic $MnPS_3$ down to monolayer thickness. Our work not only suggests $MnPS_3$ an exciting 2D-material platform for investigating novel light-magnetic ordering interactions but also shows that defect-mediated states could be utilized to reveal intrinsic magnetic properties of 2D materials.

**Materials and Methods**

*Theoretical method*: Density-functional theory (DFT) calculations were performed using the generalized gradient approximation (GGA) of Perdew-Burke-Ernzerhof (PBE) as implemented in the Vienna Ab initio Simulation Package (VASP).[56] The vdW-D2 functional correction was used to describe the long-range vdW interaction.[57] The PBE+U (U = 5 eV) approach was employed to treat the transition metal, Mn 3d electrons.[45] The cutoff energy of 500 eV was set for the plane wave expansion. The convergence of energy was set to $10^{-6}$ eV and the force on each atom was less than 0.01 eV Å$^{-1}$. The vacuum layer height along the z-direction was set to be greater than 15 Å to avoid interaction between two adjacent images. The $10 \times 10 \times 1$, $10 \times 10 \times 8$ and $4 \times 4 \times 1$ Γ-centered Monkhorst-Pack k-point grids in the first Brillouin-zone were used for primitive cell of $MnPS_3$, bulk $MnPS_3$ and $2 \times 2 \times 1$ supercell of $MnPS_3$, respectively.

*PL sample preparation*: Few-layer $MnPS_3$ samples were prepared on gold-covered 300 nm-$SiO_2$/Si substrates through standard mechanical exfoliation method.[7] Here, the 20 nm-thick gold film with the average roughness of ~ 1.78 nm on $SiO_2$/Si substrates is to prevent the background PL signal from silicon (~ 997 nm) which will disturb the PL signal of $MnPS_3$ (see **Supplementary Figure S19 and S20**). In order to prevent $MnPS_3$ samples from quality degradation, the whole exfoliation process was carried out in a $N_2$-filled commercialized glovebox with $O_2$ and $H_2O$ concentration smaller than 0.01 ppm. The samples are exposed to air within 1 minute during the transfer process from glove box to optical stage. The thickness of $MnPS_3$ flakes was determined via optical contrasts and atomic force microscopy (see Supplementary Figure S1). For UV treatment, $MnPS_3$ flakes were treated in a homemade UVO equipment. The power of the UV radiation is 100 mW.

*STEM sample preparation and characterizations*: The STEM sample was prepared using the standard polymethyl methacrylate (PMMA) assisted transfer method. Firstly, the $MnPS_3$ few-layers were exfoliated onto the $SiO_2$/Si substrate and coated with PMMA film. Then the substrate was floated on the 1 mol L$^{-1}$ KOH solution until the $SiO_2$ layer was etched away and the PMMA film with MnPS$_3$ few-layers detached from the substrate. After that, the sample was picked up by a TEM grid and rinsed in DI water several times. Finally, PMMA was removed by dipping the grid in acetone for 2 hours. The as-prepared MnPS$_3$ few-layers were further characterized using JEOL ARM-200F equipped with a CEOS CESCOR probe aberration corrector. The ADF-STEM images were collected at the accelerating voltage of 80 kV with the convergent angle of about 28 mrad and collection detector angle of 68-270 mrad.

*Optical characterizations*: We performed PL measurement in a He-flow closed-cycle cryostat (Advanced Research System) under a high vacuum (~ 10$^{-6}$ Torr) with temperatures ranging from 10 to 300 K. 405, 532 and 638 nm lasers were used as excitation sources, and were focused by a 50 × microscope objective lens (0.5 N.A.). The diameter of laser spot was ~ 3 μm for 532 nm laser. The signal was dispersed by a Andor SR-500i-D2 spectrometer with a 150 grooves/mm grating and was then detected by a CCD cooled by a thermoelectric cooler. As for Raman measurement, a 532 nm laser with power of 200 μW was used as excitation source and 600 grooves/mm grating was used for better resolution.

**Supporting Information**

Supporting Information is available from the Wiley Online Library or from the author.

**Acknowledgements**

Y. Zhou, K. He, H. Hu and G. Ouyang contributed equally to this work. We acknowledge the fruitful discussion and support from Prof. Xiaomu Wang in Nanjing University. The work was financially supported by the National Key R&D Program of China (Grant No. 2020YFA0308900), National Natural Science Foundation of China (Grant No. 61904077, 92064010, 61801210, 91833302 and U2001215), Natural Science Foundation of Jiangsu Province (Grant No. BK20180686), the funding for "Distinguished professors" and "High-level talents in six industries" of Jiangsu Province (Grant No. XYDXX-021), the Fundamental Research Funds for the Central Universities, Key Research and Development Program of Shaanxi Province (2020GXLH-Z-020 and 2020GXLH-Z-027), National Research Foundation-Competitive Research Program (NRF-CRP21-2019-0007 and NRF-CRP21-2018-0007), and the Singapore Ministry of Education Tier 3 Programme "Geometrical Quantum Materials" (MOE2018-T3-1-002).

**Contributions**

X.C. and L.W. conceived and supervised the projects. Y.Z. prepared monolayer and few-layer $MnPS_3$ on substrates. Y.Z. and X.C. performed low-temperature PL characterizations. H.H. and G.O. did the theoretical modeling. C.Z. performed STEM characterization and EDS analysis of $MnPS_3$. K.H. performed reflection spectra measurement and UVO treament of $MnPS_3$. K.H., Y.T. and R.C. performed XPS characterizations. W.W. and L.W. prepared the single-crystals of $MnPS_3$. X.C., L.W., T.W. and Y.Z. drafted the manuscript. All authors discussed and commented the manuscript.

**Conflict of interests**

The authors declare no competing financial interests.